\documentclass[12pt]{article}
\usepackage{amsmath}
\usepackage{amssymb}
\usepackage{theorem}
\usepackage{color}
\usepackage{epic}
\newcommand{\R}{{\cal R}}
\newcommand{\B}{{\cal B}}

\newcommand{\N}{{\cal N}}
\newcommand{\Nb}{{\mathbb{N}}}

\newcommand{\qed}{\hfill\hbox{\rule{6pt}{6pt}}}
\newtheorem{theorem}{Theorem}

\newtheorem{prop}{Proposition}

\newtheorem{conj}{Conjecture}
{\theorembodyfont{\rmfamily}
{\theorembodyfont{\rmfamily}\newtheorem{example}{Example}

\begin{document}

\title{Algorithmic randomness and stochastic selection function}
\author{Hayato Takahashi
\thanks{Gifu Universiy, Japan.  Email: hayato.takahashi@ieee.org}}

\maketitle

\begin{abstract}
We show algorithmic randomness versions of the two classical theorems on subsequences of normal numbers.
One is Kamae-Weiss theorem (Kamae 1973) on normal numbers, which characterize the selection function that preserves normal numbers.
Another one is the Steinhaus (1922) theorem on normal numbers, which characterize the normality from their subsequences. 
In van Lambalgen (1987), an algorithmic analogy to Kamae-Weiss theorem is conjectured in terms of algorithmic randomness and complexity.
In this paper we consider two types of algorithmic random sequence; one is ML-random sequences and the other  one is the set of sequences that have maximal complexity rate.
Then we show algorithmic randomness versions of  corresponding theorems to the above classical results.
\end{abstract}

\section{Introduction}\label{sec-intro}
Von Mises \cite{mises} seemed to try to construct a probability theory that depends on given sample sequence but does not assume probability model a priori. 
In other words von Mises tried to construct a probability theory from a statistical point of view. 
In order to achieve this program, he  introduced the notion of collective (random numbers) and demand that its  subsequences (selected with place-selection rule) have the same frequency of symbols of the original sequence. 
Since then many authors studied the properties of  subsequences of random numbers, e.g., Wald (1937), Church \cite{church}, Ville (1939).
In this paper we show algorithmic randomness versions of the two classical theorems on subsequences of normal numbers.
One is Kamae-Weiss (KW) theorem on normal numbers \cite{kamae73}, which characterize the selection function that preserves normal numbers.
Another one is the Steinhaus theorem on normal numbers \cite{steinhaus22}, which characterize the normality from their subsequences. 
In van Lambalgen \cite{lambalgen87}, an algorithmic analogy to KW theorem is conjectured in terms of algorithmic randomness and complexity \cite{chaitin75,Kol65,martin-lof66,solomonoff64}. 
In this paper we consider two types of algorithmic random sequence; one is Martin-L{\"o}f (ML)-random sequences and the other  one is the set of sequences that have maximal complexity rate.
Then we show algorithmic randomness versions of  corresponding theorems to the above classical results.

Let \(\Omega\) be the set of infinite binary sequences.
For \(x,y\in\Omega\), let \(x=x_1x_2\cdots\) and \(y=y_1y_2\cdots,\ \forall i\ x_i, y_i\in\{0,1\}\).
Let \(\tau:\Nb\to\Nb\) be a strictly increasing function such that 
\(\{ i\mid y_i=1\}=\{\tau(1)<\tau(2)<\cdots\}\), where \(\Nb\) is the set of natural numbers. 
If \(\sum_i y_i=n\) then \(\tau(j)\) is defined for \(1\leq j\leq n\).
For \(x,y\in\Omega\) let \(x/y\) be the subsequence of \(x\) selected at \(y_i=1\), i.e.,
\(x/y=x_{\tau(1)}x_{\tau(2)}\cdots\).
For example, if \(x=0011\cdots,\ y=0101\cdots\) then \(\tau(1)=2, \tau(2)=4\) and \(x/y=01\cdots\).
For finite binary strings  \(x_1^n:=x_1\cdots x_n\) and \(y_1^n:=y_1\cdots y_n\),
\(x_1^n/y_1^n\) is defined similarly. 
Let \(S\) be the set of finite binary strings and \(|s|\) be the length of \(s\in S\).
For \(s\in S\) let \(\Delta(s):=\{s\omega | \omega\in\Omega\}\), where \(s\omega\) is the concatenation of \(s\) and \(\omega\).
Let \((\Omega,\B, P)\) be a probability space, where \(\B\) is the sigma-algebra generated by \(\Delta(s), s\in S\).
We write \(P(s):=P(\Delta(s))\).

A probability \(P\) on \(\Omega\) is called computable if there is a computable function \(A\) such that \(\forall s, k\ |P(s)-A(s,k)|<1/k\).
For \(A\subset S\), let \(\tilde{A}:=\cup_{s\in A}\Delta(s)\).
A recursively enumerable (r.e.) set \(U\subset \Nb\times S\) is called (ML) test with respect to \(P\) if 1) \(U\) is r.e., 2) \(\tilde{U}_{n+1}\subset \tilde{U}_n\) for all \(n\), where \(U_n=\{s : (n,s)\in U\}\), and
3) \(P(\tilde{U}_n)<2^{-n}\). 
A test \(U\) is called universal if for any other test \(V\), there is a constant \(c\) such that \(\forall n\ \tilde{V}_{n+c}\subset \tilde{U}_n\).
In \cite{martin-lof66}, it is shown that a universal test \(U\) exists if \(P\) is computable and the set \(\R^P:=(\cap_{n=1}^\infty \tilde{U}_n)^c\) is called the set of ML-random sequences with respect to \(P\).

Next, we introduce another notion of randomness. 
We say that \(y\) has maximal complexity rate with respect to \(P\) if 
\begin{equation}\label{weak}
\lim_{n\to\infty}\frac{1}{n} K(y_1^n)=\lim_{n\to\infty}-\frac{1}{n}\log P(y_1^n),
\end{equation}
i.e., both sides exist and are equal. 
For example, \(y\) has maximal complexity rate with respect to the uniform measure (i.e., \(P(s)=2^{-|s|}\) for all \(s\)) if\\
  \(\lim_{n\to\infty} K(y_1^n)/n=1\).
If \(y\) is ML-random sequences with respect to a computable ergodic \(P\)  then
from upcrossing inequality for the Shannon-McMillan-Breiman theorem \cite{hochman2009}, the right-hand-side of (\ref{weak}) exists (see also \cite{vyugin98}) and from Levin-Schnorr theorem (see 
(\ref{ls}) below), we see that (\ref{weak}) holds i.e., \(y\) has maximal complexity rate w.r.t. \(P\).

\section{Algorithmic version of Kamae-Weiss theorem}

In Kamae \cite{kamae73}, it is shown that the following two statements are equivalent under the assumption that \(\liminf \frac{1}{n}\sum_{i=1}^n y_i>0\):
\begin{theorem}[Kamae-Weiss]\hfill\\
(i) \(h(y)=0\).\\
(ii) \(\forall x\ x\in \N\to x/y\in \N\),\\
where \(h(y)\) is  Kamae entropy \cite{brudno83,lambalgen87} and \(\N\) is the set of binary normal numbers.
\end{theorem}
A probability \(p\) on \(\Omega\) is called cluster point if there is a sequence \(\{n_i\}\)
\[\forall s\in S\ p(s)=\lim_{i\to \infty} \#\{ 1\leq j\leq n_i \mid x_j\cdots x_{j+|s|-1}=s\}/n_i.\]
From the definition,  the cluster points are stationary measures. 
Let \(V(x)\) be the set of cluster points defined from \(x\).
 From a diagonal argument we see that \(V(x)\ne\emptyset\) for all \(x\).
Kamae entropy is defined by
\[h(x)=\sup\{ h(p)\mid p\in V(x)\},\]
where \(h(p)\) is the measure theoretic entropy of \(p\).
If \(h(x)=0\), it is called completely deterministic, see \cite{kamae73,weiss71,weiss00}.
The part (i)\(\Rightarrow\) (ii) is appeared in \cite{weiss71}.

As a natural analogy,  the following equivalence (algorithmic randomness version of Kamae's theorem) under a suitable restriction on \(y\) is conjectured   in van Lambalgen \cite{lambalgen87},
\begin{conj}[Lambalgen]\hfill\\
(i) \(\lim_{n\to\infty}K(y_1^n)/n=0.\)\\
(ii) \(\forall x\ x\in\R\to x/y\in\R\)
\end{conj}
where \(K\) is the prefix Kolmogorov complexity and \(\R\) is the set of ML-random sequences with respect to the uniform measure (fair coin flipping), see \cite{LV2008}.
Note that  \(\lim_{n\to\infty}K(y_1^n)/n=h, P-a.s.,\) for ergodic \(P\) and its entropy \(h\), see \cite{brudno83}.
%\section{Results}
We show two algorithmic analogies to KW theorem (the following results are appeared in Takahashi \cite{takahashi2011} however we reproduce them for convenience).
The first one is a ML-randomness analogy and the second one is a complexity rate analogy to KW theorem, respectively.

Our first algorithmic analogy to the KW theorem is the following. 
\begin{prop}[\cite{takahashi2011}]
Suppose that \(y\) is ML-random with respect to some computable probability \(P\) and \(\sum_{i=1}^\infty y_i=\infty\).
Then the following two statements are equivalent:\\
(i) \(y\) is computable.\\
(ii) \(\forall x\ x\in\R\to x/y\in\R^y\),\\
where \(\R^y\) is the set of ML-random sequences with respect to the uniform measure relative to \(y\). 
\end{prop}
Proof) (i)\(\Rightarrow\) (ii).
Since \(\sum_{i=1}^\infty y_i=\infty\) we have   \(\forall s\ \lambda \{ x\in\Omega\mid s\sqsubset x/y\} =2^{-|s|}\), where \(\lambda\) is the uniform measure.
Let \(U\) be a universal test with respect to \(\lambda\) and \(y(s)\subset S\) be a finite set such that \(\{ x\in\Omega\mid s\sqsubset x/y\}= \tilde{y}(s)\).
Then \(y(s)\) is computable from \(y\) and \(s\), and hence \(U^y:=\{ (n, a)\mid a\in y(s), s\in U_n\}\) is a test if \(y\) is  computable.
We have \(x\in \tilde{U}_n^y\leftrightarrow x/y\in \tilde{U}_n\).
(Intuitively  \(U^y\) is a universal test on subsequences selected by \(y\)).
Then 
\[
x\in\R  \to x\notin \cap_n\tilde{U}_n^y \leftrightarrow x/y \notin \cap_n\tilde{U}_n \leftrightarrow x/y\in\R.
\]
Since \(y\) is computable, \(\R^y=\R\) and we have (ii).

Conversely,  suppose  that \(y\) is a  ML-random sequence with respect to a computable \(P\) and is not computable. 
From Levin-Schnorr theorem, we have
\begin{equation}\label{ls}
\forall n\ Km(y_1^n)=-\log P(y_1^n)+O(1),
\end{equation}
where \(Km\) is the monotone complexity. 
Throughout the paper, the base of logarithm is 2.
By applying arithmetic coding to \(P\), there is a sequence \(z\) such that \(z\) is computable from \(y\) and
\(y^n_1\sqsubset u(z^{l_n}_1),\ l_n= -\log P(y^n_1)+O(1)\) for all \(n\), where \(u\) is a monotone function and we write \(s\sqsubset s'\) if \(s\) is a prefix of \(s'\).
Since \(y\) is not computable, we have \(\lim_n l_n=\infty\). 
From (\ref{ls}), we see that \(\forall n\ Km(z_1^{l_n})=l_n+O(1)\).
We show that if \(y\in\R^P\) then \(\sup_n  l_{n+1}-l_n  <\infty\).
Observe that if \(y\in\R^P\) then \(\forall n\ P(y_1^n)>0\) and 
\begin{align*}
\sup_n   l_{n+1}-l_n  <\infty\leftrightarrow \sup_n -\log P(y_{n+1}\mid y_1^n)<\infty & \leftrightarrow \inf_n P(y_{n+1}\mid y^n_1)>0\\
& \leftrightarrow \liminf_n P(y_{n+1}\mid y^n_1)>0.
\end{align*}
Let  \(U_n:=\{ s \mid P(s \mid s_1^{|s|-1})<2^{-n}\}\). Then \(P(\tilde{U}_n)<2^{-n}\) and  \(U:=\{ (n, s)\mid s\in U_n\}\) is a r.e.~set. 
Since \(y\in\limsup_n \tilde{U}_n\leftrightarrow \liminf_n P(y_{n+1}\mid y_1^n)=0\), we see that  if \(y\in\R^P\) then \(\sup_n   l_{n+1}-l_n  <\infty\).
(If \(U\) is r.e.~and \(P(\tilde{U}_n)<2^{-n}\) then \(\R^P\subset (\limsup_n \tilde{U}_n)^c\), see \cite{shen89}.)
Since  \(\forall n\ Km(z_1^{l_n})=l_n+O(1)\) and \(\sup_n   l_{n+1}-l_n  <\infty\), we have \(\forall n\ Km(z_1^n)=n+O(1)\) and \(z\in\R\).
Since \(z\) is computable from \(y\) we have \(z/y\notin \R^y\).
\qed

\begin{prop}[\cite{takahashi2011}]\label{main}
Suppose that \(y\) has maximal complexity rate with respect to a computable probability and \(\lim_n \frac{1}{n}\sum_{i=1}^n y_i>0\).
Then the following two statements are equivalent:\\
(i) \(\lim_{n\to\infty}\frac{1}{n}K(y_1^n)=0.\)\\
(ii) \(\forall x\ \lim_{n\to\infty}\frac{1}{n}K(x_1^n)=1\to \lim_{n\to\infty} \frac{1}{|x_1^n/y_1^n|}K(x_1^n/y_1^n |y_1^n )=1.\)
\end{prop}
Proof)\\
(i) \(\Rightarrow\) (ii)\\
Let \(\bar{y}:=\bar{y}_1\bar{y}_2\cdots\in\Omega\) such that \(\bar{y}_i=1\) if \(y_i=0\) and \(\bar{y}_i=0\) else for all \(i\).
Since 
\[ | K(x_1^n)-K(x_1^n | y_1^n) | \leq K(y_1^n)+O(1)\text{ and }\]
 \begin{align*}
& K(x_1^n | y_1^n)=K(x_1^n/ y_1^n, x_1^n/ \bar{y}_1^n | y_1^n) +O(1),
\end{align*}
if \(\lim_{n\to\infty}K(y_1^n)/n=0\) and \(0<\lim_n \frac{1}{n}\sum_{i=1}^n y_i<1\) then 
we have
\begin{align*}
& \lim_{n\to\infty}K(x_1^n)/n=1\\
& \Rightarrow \lim_{n\to\infty} \frac{1}{n} K(x_1^n/ y_1^n, x_1^n/ \bar{y}_1^n | y_1^n)=1\\
& \Rightarrow \lim_{n\to\infty} \frac{1}{n}(K(x_1^n/ y_1^n| y_1^n)+K(x_1^n/ \bar{y}_1^n | y_1^n))=1\\
& \Rightarrow \lim_{n\to\infty} \frac{n_1}{n}\frac{1}{n_1}K(x_1^n/ y_1^n| y_1^n)+\frac{n-n_1}{n}\frac{1}{n-n_1}K(x_1^n/ \bar{y}_1^n | y_1^n)=1\\
& \Rightarrow \lim_{n\to\infty} \frac{1}{n_1}K(x_1^n/y_1^n |y_1^n )=1\text{ and }\lim_{n\to\infty}\frac{1}{n-n_1}K(x_1^n/ \bar{y}_1^n | y_1^n)=1.
\end{align*}
where \(n_1=|x_1^n/y_1^n|=\sum_{i=1}^n y_i\).
Similarly, if \(\lim_{n\to\infty}K(y_1^n)/n=0\)  and \\
\(\lim_n \frac{1}{n}\sum_{i=1}^n y_i=1\) then we have 
\( \lim_{n\to\infty} \frac{1}{n_1}K(x_1^n/y_1^n |y_1^n )=1\).\\
(ii) \(\Rightarrow\) (i)\\
Suppose that 
\begin{equation}\label{wls}
\lim_{n\to\infty}\frac{1}{n} K(y_1^n)=\lim_{n\to\infty}-\frac{1}{n}\log P(y_1^n)>0,
\end{equation}
for a computable \(P\).
Let \(l_n\) be the least integer greater than \(-\log P(y_1^n)\).
Then by considering arithmetic coding, there is \(z=z_1z_2\cdots\in\Omega\) and a monotone function \(u\)  such that 
\(y^n_1\sqsubset u(z^{l_n}_1)\).  By considering optimal code for \(z^{l_n}_1\) we have \(Km(y^n_1)\leq Km(z^{l_n}_1)+O(1)\).
From  (\ref{wls}), we have \(\lim_n Km(y^n_1)/l_n=\lim_n  Km(z^{l_n}_1)/l_n=1\).
For \(l_n\leq t \leq l_{n+1}\), we have \(Km(z^{l_n}_1)/l_{n+1}\leq Km(z^t_1)/t\leq Km(z^{l_{n+1}}_1)/l_n\).
From (\ref{wls}), we have \(\lim_n l_{n+1}/l_n=1\), and hence \(\lim_n Km(z^n_1)/n=\lim_n K(z^n_1)/n=1\).

Since 1) \(z^{l_n}_1\) is computable from \(y^n_1\), 2)  \(\lim_n l_n/n>0\) by (\ref{wls}), and \\
3)  \(\lim_n \frac{1}{n}\sum_{i=1}^n y_i>0\),  we have \(\limsup_{n\to\infty} \frac{1}{|z_1^n/y_1^n|}K(z_1^n/y_1^n | y_1^n)<1\).
\qed

\begin{example}
Champernowne sequence satisfies the condition of the proposition and (i) holds, however its Kamae-entropy is not zero. 
\end{example}

\begin{example}
If \(y\) is a Sturmian sequence  generated by an irrational rotation model with a computable parameter \cite{kamaeTakahashi,takahashiAndaihara} then \(y\) satisfies the condition of the proposition and (i) holds. 
\end{example}

\section{Algorithmic  version of Steinhaus theorem}
In Steinhaus  \cite{steinhaus22},  it is shown that 
\begin{theorem}[Steinhaus]
\(x\) is normal number iff for all \(q\in (0,1),\ u_q\{ y \mid  x/y \text { is normal number}\}=1\), where \(u_q\) is the binary i.i.d. process with parameter
\(q\), i.e., \(u_q(1)=q, u_q(0)=1-q\).
\end{theorem}

We have an algorithmic analogies for this result.

\begin{prop}\label{model-selection}
Let \(q\in [0,1]\). The following two statements are equivalent:\\
(i) \(x\in\R^{u_q, q}\)\\
(ii) \(\exists\text{~computable }y,\ x\in\cup_{r\in [0,1]}\R^{u_r, r}\text{ and } x/y\in \R^{u_q, q}\),\\
where \(\R^{u_q, q}\) is the set of ML-random sequences w.r.t.~\(u_q\) relative to \(q\).
\end{prop}
Proof) 
By considering the ML-test on the subseqences selected by \(y\), (i)\(\Rightarrow\) (ii) follows. 
Conversely if \(x/y\in \R^{u_q, q}\) then \(x/y\) satisfies the law of large numbers, for example, see \cite{LV2008}. 
Thus \(q\) is uniquely determined (in fact computable) from \(x/y\) and we have (i).
\qed

\begin{prop}
Let \(w\) be a computable probability such that\\
(a) \(\forall y\in\R^w,\ \lim_n K(y^n)/n=0\), (b) \(\lim_n \sum_{1\leq i\leq n} y_i/n\) exists for \(y\in\R^w\), and 
(c) \(\forall \epsilon >0 \exists y\in\R^w  \lim_n \sum_{1\leq i\leq n} y_i/n>1-\epsilon\).\\
Then the following two statements are equivalent.\\
(i) \(\lim_{n\to\infty} \frac{1}{n}K(x^n)=1\).\\
(ii) \(\lim_{n\to\infty} \frac{1}{| x^n/y^n|}K(x^n/y^n)=1\) for \(y\in\R^w\).
\end{prop}
Proof)
(i) \(\Rightarrow\) (ii) follows from the part of (i)\(\Rightarrow\) (ii) of Proposition~2.\\
(ii)\(\Rightarrow\) (i): Observe that
\begin{gather}
\vert K(x^n, y^n)-K(x^n)\vert \leq K(y^n)+O(1),\label{stein-A}\\
\vert K(x^n,y^n)-K(x^n/y^n, x^n/ \bar{y}^n, y^n)\vert \leq O(1),\label{stein-B}\\
\vert K(x^n/y^n)- K(x^n/y^n, x^n/ \bar{y}^n, y^n)\vert \leq K(x^n/\bar{y}^n)+K(y^n)+O(1), \label{stein-C}
\end{gather}
where \(\bar{y}\) is defined in the proof of Proposition~\ref{main}.
From the condition, we have
\(\forall \epsilon >0\ \exists y\in\R^w\ 1-\epsilon <\lim_n \frac{\vert x^n/y^n\vert}{n}\leq 1\).
Then  \(\forall \epsilon >0\ \exists y\in\R^w \limsup_n \frac{1}{\vert x^n/y^n\vert}K(x^x/\bar{y}^n)=\limsup_n \frac{\vert x^n/\bar{y}^n\vert}{\vert x^n/y^n\vert}\frac{1}{\vert x^n/\bar{y}^n\vert}K(x^n/\bar{y}^n)\leq\frac{\epsilon}{1-\epsilon}\) \\
and  \(\limsup_n \frac{1}{\vert x^n/y^n\vert} K(y^n)=0\), where the latter equality follows from the condition (a).
From (ii), (\ref{stein-A}), (\ref{stein-B}), and (\ref{stein-C}), we have 
\(\forall \epsilon >0\ \exists y\in\R^w 1-\frac{\epsilon}{1-\epsilon}\leq \liminf_n\frac{1}{\vert x^n/y^n\vert}K(x^n)\leq \limsup_n \frac{1}{\vert x^n/y^n\vert}K(x^n)\leq 1+\frac{\epsilon}{1-\epsilon}\).
Since we can choose \(\epsilon>0\) arbitrary, we have (i).
\qed\\
\begin{example}
Let \(w:=\int P_\rho d\rho\), where \(P_\rho\) is a probability derived from irrational rotation with parameter \(\rho\).
Then \(w\) satisfies the condition of Prop.~4, see\cite{takahashiAndaihara}.
\end{example}

\section{Discussion}
Both proofs of Proposition 1 and 2 have similar structure, i.e., the part (i) \(\to\) (ii) are straightforward
 and in order to show the converse, 
we construct random sequences (in the sense of Proposition 1 and 2, respectively) by compression. 

We may say that Proposition 1 is a Martin-L\"of randomness analogy and Proposition 2 is a complexity rate analogy to KW theorem, respectively. 
These results neither prove nor disprove the conjecture of van Lambalgen.
However Martin-L\"of randomness and complexity rate randomness give different  classes of randomness, and 
a strange point of the conjecture is that it is described in terms of different notions of randomness.

\begin{center}
{\bf Acknowledgement}
\end{center}
The author thanks Prof.~Teturo Kamae (Matsuyama Univ.) for discussions and comments.


\begin{thebibliography}{10}

\bibitem{brudno83}
A.~A. Brudno.
\newblock Entropy and the complexity of the trajectories of a dynamical system.
\newblock {\em Trans.~Mosc.~Math.~Soc.}, 44:127--151, 1983.

\bibitem{chaitin75}
G.~J. Chaitin.
\newblock A theory of program size formally identical to information theory.
\newblock {\em J.~ACM}, 22:329--340, 1975.

\bibitem{church}
A.~Church.
\newblock On the concept of a random sequence.
\newblock {\em Bull.~Amer.~Math.~Soc.}, 46:130--135, 1940.

\bibitem{hochman2009}
M.~Hochman.
\newblock Upcrossing inequalities for stationary sequences and applications.
\newblock {\em Ann.~Probab.}, 37(6):2135--2149, 2009.

\bibitem{kamae73}
T.~Kamae.
\newblock Subsequences of normal numbers.
\newblock {\em Israel J.~Math.}, 16:121--149, 1973.

\bibitem{kamaeTakahashi}
T.~Kamae and H.~Takahashi.
\newblock Statistical problems related to irrational rotations.
\newblock {\em Ann.~Inst.~Statist.~Math.}, 58(3):573--593, 2006.

\bibitem{Kol65}
A.~N. Kolmogorov.
\newblock Three approaches to the quantitative definition of information.
\newblock {\em Probl.~Inf.~Transm.}, 1(1):1--7, 1965.

\bibitem{LV2008}
M.~Li and P.~Vit{\'a}nyi.
\newblock {\em An introduction to Kolmogorov complexity and Its applications}.
\newblock Springer, New York, third edition, 2008.

\bibitem{martin-lof66}
P.~Martin-L{\"o}f.
\newblock The definition of random sequences.
\newblock {\em Information and Control}, 9:602--609, 1966.

\bibitem{shen89}
A.~Kh. Shen.
\newblock On relations between different algorithmic definitions of randomness.
\newblock {\em Soviet Math.~ Dokl.}, 38(2):316--319, 1989.

\bibitem{solomonoff64}
R.~J. Solomonoff.
\newblock A formal theory of inductive inference, part 1 and part2.
\newblock {\em Inform.~Contr.}, 7:1--22, 224--254, 1964.

\bibitem{steinhaus22}
H.~Steinhaus.
\newblock Les probabilit{\'e}s d{\'e}nombrables et leur rapport {\`a} la
  th{\'e}orie de la me{\'e}sure.
\newblock {\em Fund.~Math.}, 4:286--310, 1922.

\bibitem{takahashi2011}
H.~Takahashi.
\newblock Algorithmic analogies to kamae-weiss theorem on normal numbers.
\newblock In {\em Solomonoff 85th memorial conference}, 2011.
\newblock To appear in LNAI.

\bibitem{takahashiAndaihara}
H.~Takahashi and K.~Aihara.
\newblock Algorithmic analysis of irrational rotations in a sigle neuron model.
\newblock {\em J.~Complexity}, 19:132--152, 2003.

\bibitem{lambalgen87}
M.~van Lambalgen.
\newblock {\em Random sequences}.
\newblock PhD thesis, Universiteit van Amsterdam, 1987.

\bibitem{mises}
R.~von Mises.
\newblock {\em Probability, Statistics and Truth}.
\newblock Dover, 1981.

\bibitem{vyugin98}
V.~V. V'yugin.
\newblock Ergodic theorems for individual random sequences.
\newblock {\em Theor.~Comp.~Sci.}, 207:343--361, 1998.

\bibitem{weiss71}
B.~Weiss.
\newblock Normal sequences as collectives.
\newblock In {\em Proc.~Symp.~on Topological Dynamics and Ergodic Theory}.
  Univ.~of Kentucky, 1971.

\bibitem{weiss00}
B.~Weiss.
\newblock {\em Single Orbit Dynamics}.
\newblock Amer.~Math.~Soc., 2000.

\end{thebibliography}
\end{document}